\documentclass{wsm}
\usepackage{amsfonts,amssymb,graphicx,bm}
\newcommand{\LQCD}{\Lambda_{\mbox{\tiny QCD}}}
\newcommand{\QbQ}{$\mbox{Q}\overline{\mbox{Q}}$ }
\newcommand{\nf}{n_{\mbox{\scriptsize f}}}
\newcommand{\nfs}{n_{\mbox{\tiny f}}}
\newcommand{\Reg}{{\mbox{\scriptsize Reg}}}

\begin{document}

\markboth{}{}

%
\catchline{}{}{}{}{}
%

\title{ANALYTIC INVARIANT CHARGE AND THE LATTICE STATIC
QUARK--ANTIQUARK POTENTIAL}

\author{\footnotesize A.V.~NESTERENKO}

\address{Centre de Physique Th\'eorique de l'\'Ecole Polytechnique \\
         91128 Palaiseau Cedex, France\footnote{Unit\'e Mixte de
         Recherche du CNRS (UMR 7644)} \\
         nesterav@cpht.polytechnique.fr}

\address{Bogoliubov Laboratory of Theoretical Physics,
         Joint Institute for Nuclear Research \\
         Dubna 141980, Russian Federation \\
         nesterav@thsun1.jinr.ru}

\maketitle

\pub{Received 12 December 2003}{}

\begin{abstract}
A recently developed model for the QCD analytic invariant charge is
compared with quenched lattice simulation data on the static
quark--antiquark potential.  By employing this strong running
coupling one is able to obtain the confining quark--antiquark
potential in the framework of the one--gluon exchange model.  To
achieve this objective a technique for evaluating the integrals of a
required form is developed. Special attention is paid here to
removing the divergences encountered the calculations. All this
enables one to examine the asymptotic behavior of the potential at
both small and large distances with high accuracy. An explicit
expression for the quark--antiquark potential, which interpolates
between these asymptotics, and satisfies the concavity condition, is
proposed. The derived potential coincides with the perturbative
results at small  distances, and it is in a good agreement with the
lattice data in the nonperturbative physically--relevant region. An
estimation of the parameter $\LQCD$ is obtained for the case of pure
gluodynamics. It is found to be consistent with all the previous
estimations of $\LQCD$ in the framework of approach in hand.

\end{abstract}


\section{Introduction}

     The description of hadron dynamics at large distances remains a
crucial challenge of elementary particle physics for a long time. On
the one hand, the asymptotic freedom of Quantum Chromodynamics (QCD)
allows one to apply perturbation theory to study some
``short--range'' phenomena, for instance, the lowest--lying bound
states of heavy quark--antiquark systems. On the other hand, for the
consistent description of many phenomena related to the
``long--range'' dynamics (such as confinement of quarks, structure of
the QCD vacuum, etc.) something more than the usual perturbative
approach has to be involved.

     Theoretical analysis of the strong interaction basically relies
upon the renormalization group (RG) method. Usually, in order to
describe the hadron dynamics in the asymptotical ultraviolet (UV)
region, one applies the RG method together  with perturbative
calculations. However, this leads to unphysical singularities of the
outcoming solutions to RG equations, that contradicts the general
principles of local Quantum Field Theory (QFT).  An effective way to
overcome such difficulties consists  in invoking the analyticity
requirement. This prescription became the underlying idea of  the
so-called analytic approach to QFT,\cite{Redmond,BLS} which was
lately extended to QCD by Shirkov and Solovtsov.\cite{SS97,DV01}

     In the framework of the analytic approach to QCD a new  model
for the strong running coupling has recently been
developed.\cite{PRD1,PRD2} Its basic idea is to impose the
analyticity requirement on the RG $\beta$ function perturbative
expansion for restoring its correct analytic properties (see Sec.~2
for the details). The analytic invariant charge possesses a number of
profitable properties,  in particular, it contains no unphysical
singularities (see also Refs.~\refcite{IJMPA,MPLA1,MPLA2} for the
description of its traits). It is worth noting here that this
analytic running coupling incorporates the perturbative and
intrinsically nonperturbative features of the strong interaction, and
enables one to describe a wide range of hadron processes.

     A crucial insight into the nonperturbative aspects of the strong
interaction at large distances can be provided by lattice
simulations. This may be, for instance, the static  quark--antiquark
($\mbox{Q}\overline{\mbox{Q}}$) potential calculated up to the
significantly large distances\cite{Bali2000} ($r \gtrsim 1\,$fm), or
the investigation of the topological structure of the QCD
vacuum.\cite{UKQCD} Certainly, a decisive test of any model for the
strong interaction is its comparison with the lattice results.

     The objective of this paper is to study the asymptotic behavior
of the static quark--antiquark potential constructed by making use of
the analytic invariant charge. It is also of  a primary interest to
collate the derived \QbQ potential with the relevant lattice
simulation data. For verification the consistency of the results it
is worth to evaluate the  corresponding parameter $\LQCD$ and to
compare it with its previous estimations obtained in the framework
of approach in hand.

     The layout of the paper is as follows. In Sec.~2 the model for
the QCD analytic invariant charge is briefly described  and its basic
features are outlined. In Sec.~3 the asymptotic behavior of the
quark--antiquark potential, constructed by making use of  this strong
running coupling, is examined. A technique for treating the divergent
integrals encountered is developed here. The explicit expression for
the \QbQ potential, which possesses the obtained asymptotics, and
satisfies the concavity condition, is derived. In Sec.~4 these
results are applied to study of the relevant lattice simulation data.
The constructed quark--antiquark potential agrees fairly well with
the lattice data in the nonperturbative physically--relevant region.
At the  same time, it coincides with the perturbative \QbQ potential
at  small distances. The fitted parameter $\LQCD$ is found to be in a
good agreement with all its previous estimations, implying the
consistency of obtained results. In the Conclusion (Sec.~5) the
achieved goals are formulated in a compact way.

\section{The QCD Analytic Invariant Charge}

\begin{figure}[ht]
\centerline{\includegraphics[width=65mm]{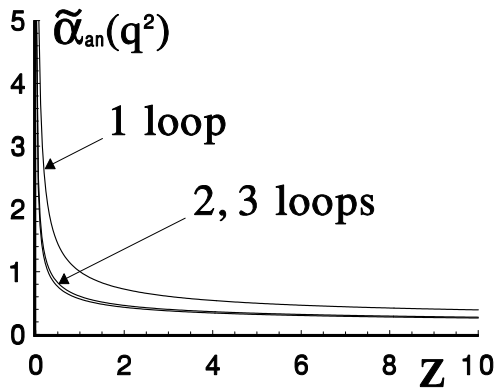}}
\caption{The QCD analytic invariant charge at different loop levels,
$z=q^2/\Lambda^2$.
\label{Fig:AIC}}
\end{figure}

     As it has been mentioned in the Introduction, the perturbative
approximation of the $\beta$~function in the RG equation for the
strong running coupling $\alpha(\mu^2) = g^2(\mu^2)/(4 \pi)$
\begin{equation}
\frac{d\, \ln\left[g^2(\mu^2)\right]}{d\,\ln\mu^2} =
\beta\left(g(\mu^2)\right)
\end{equation}
leads to unphysical singularities of the outcoming solutions, that
contradicts the general principles of local Quantum Filed  Theory.  A
plausible way to overcome such difficulties consists in imposing the
analyticity requirement on the perturbative expansion of the RG
$\beta$~function for restoring its correct analytic properties. This
prescription is a distinctive feature of the recently developed model
for the QCD analytic invariant charge.\cite{PRD1,PRD2} At the
one-loop level the corresponding renormalization group equation can
be solved explicitly:
\begin{equation}
\label{AIC}
\alpha_{\mbox{\scriptsize an}}(q^2) = \frac{4 \pi}{\beta_0}\,
\frac{z-1}{z\,\ln z}, \qquad z = \frac{q^2}{\Lambda^2},
\end{equation}
while at the higher loop levels only the integral representation for
the analytic invariant charge has been obtained (see
Refs.~\refcite{PRD2,MPLA2}). Figure~\ref{Fig:AIC} presents the
analytic running coupling $\widetilde{\alpha}_{\mbox{\scriptsize
an}}(q^2) = \alpha_{\mbox{\scriptsize an}}(q^2) \beta_0/(4 \pi)$ at
different loop levels.

     The developed model for the QCD analytic invariant charge
possesses a number of appealing traits (see
Refs.~\refcite{MPLA1,MPLA2}). Namely, it has no unphysical
singularities at any loop level; it contains no adjustable parameters;
it incorporates ultraviolet asymptotic freedom with infrared (IR)
enhancement in a single expression; it has universal behavior both in
UV and IR regions at any loop level; it possesses a good higher loop
and scheme stability. This model also enables one to describe various
strong interaction processes both of perturbative and intrinsically
nonperturbative nature (see papers~\refcite{IJMPA,QCD03,ConfV} and
references therein). It is of a primary importance to mention here
that the  analytic invariant charge~(\ref{AIC}) has recently been
rediscovered when studying the conformal inversion symmetry related
to the size distribution of instantons.\cite{Schrempp} In particular,
Eq.~(\ref{AIC}) was proved to reproduce explicitly this kind of
symmetry. In turn, the latter is in a good agreement with the
relevant lattice data by the UKQCD Collaboration\cite{UKQCD} (see
Refs.~\refcite{Schrempp} and~\refcite{IJMPA} for the details).

\section{The Static Quark--Antiquark Potential}

     Let us proceed now to the construction of the static
quark--antiquark potential. In the framework of the one-gluon
exchange model it is related to the strong running coupling
$\alpha(q^2)$ by the 3-dimensional Fourier transformation
\begin{equation}
\label{VrGen}
V(r) = - \frac{16 \pi}{3} \int\limits_{0}^{\infty}
\frac{\alpha(\bm{q}^2)}{\bm{q}^2}\,
\frac{\exp(i \bm{q r})}{(2 \pi)^3}\, d {\bm q}.
\end{equation}
Strictly speaking, this definition of the potential is justified
for small distances ($r \lesssim 0.1\,$fm) only. For example, the
lowest--lying bound states of heavy quarks can be described by
employing the perturbative\footnote{The leading short--distance
nonperturbative effect due to the gluon condensate has also been
taken into account in Ref.~\refcite{Ynd1}.} QCD.\cite{Ynd1}
However, at large  distances ($r \gtrsim 0.5\,$fm), which play the
crucial role in hadron spectroscopy, the perturbative approach blows
up due to unphysical singularities (such as the Landau pole) of the
strong running coupling. Nevertheless, the model (\ref{VrGen}), being
complemented with a certain insight into the nonperturbative behavior
of the QCD invariant charge, has proved to be successful for description
both heavy--quark and light--quark systems (see, e.g.,
reviews~\refcite{Lucha91,Brambilla99,Kiselev1} and references therein
for the details).

     In this paper, for the construction of the static potential of
the quark--antiquark interaction, we shall use the analytic invariant
charge (\ref{AIC}). After integration over angular variables,
Eq.~(\ref{VrGen}) in this case takes the form
\begin{equation}
\label{VrInt}
V(r) = - \frac{32}{3 \beta_0}\, \frac{1}{r_0} \int\limits_{0}^{\infty}
\frac{p^2-1}{p^2\, \ln p^2}\, \frac{\sin (p R)}{p R}\, d p,
\end{equation}
where $p = q r_0$, $R = r/r_0$, and $r_0$ is a reference scale of the
dimension of length, which will be specified below.  This integral
diverges at the lower limit, that is a common feature of the models
of such kind (see, e.g., Ref.~\refcite{Lucha91}). We shall treat this
divergence by employing the analytical regularization (see
Eq.~(\ref{IntCont})).

     The analysis of Eq.~(\ref{VrInt}), performed in
Ref.~\refcite{PRD1}, elucidated only the leading asymptotic behavior of
$V(r)$ at large distances. However, it is still desirable to examine
at a more precise level the asymptotics of the quark--antiquark
potential~(\ref{VrInt}) in the both, ultraviolet and infrared
regions. This objective can be achieved in the following way. First
of all, let us introduce a dimensionless variable $Q = p R$ and
rewrite Eq.~(\ref{VrInt}):
\begin{equation}
\label{VrInt1}
V(r) = \frac{16}{3 \beta_0}\, \frac{1}{r_0} \, \frac{1}{R}
\int\limits_{0}^{\infty}\!\left(1 - \frac{R^2}{Q^2}\right)
\frac{\sin Q}{Q}\, \frac{d Q}{\ln R - \ln Q}.
\end{equation}
Then, formally expanding the denominator of the integrand,\footnote{A
similar method has also been used in Ref.~\refcite{LeTo}.}  one can
find the asymptotic behavior of $V(r)$ at both small and large
distances:
\begin{eqnarray}
V(r) &\simeq& \frac{16}{3 \beta_0}\,\frac{1}{r_0}
\Biggl[
\frac{1}{R} \sum_{n=0}^{n_0}\frac{1}{(\ln R)^{n+1}}
\int\limits_{0}^{\infty} \frac{\sin Q}{Q} (\ln Q)^n\, d Q
\nonumber \nopagebreak \\ \nopagebreak
& & - R \sum_{m=0}^{m_0}\frac{1}{(\ln R)^{m+1}}
\int\limits_{0}^{\infty} \frac{\sin Q}{Q^3} (\ln Q)^m\, d Q
\Biggr].
\label{VrIntExp}
\end{eqnarray}
The values of $n_0$ and $m_0$ will be specified below.

     For evaluation of the expansion coefficients in
Eq.~(\ref{VrIntExp}) it is worth considering an  auxiliary integral
of the form:
\begin{equation}
\label{IntAux1}
\int\limits_{0}^{\infty} Q^t\, \sin Q \, d Q = \sqrt{\pi}\, 2^t \,
\frac{\Gamma\left(1+\frac{t}{2}\right)}
{\Gamma\left(\frac{1}{2} - \frac{t}{2}\right)}.
\end{equation}
Differentiating this equation $n$ times with respect to variable $t$
we obtain
\begin{equation}
\label{IntAux2}
\int\limits_{0}^{\infty} Q^t\, (\ln Q)^n \, \sin Q \, d Q = v(n, t),
\end{equation}
where
\begin{equation}
\label{vDef}
v(n, t) = \sqrt{\pi}\, \frac{d^n}{d t^n} \left[2^t\,
\frac{\Gamma\left(1+\frac{t}{2}\right)}
{\Gamma\left(\frac{1}{2} - \frac{t}{2}\right)}
\right].
\end{equation}
In Eq.~(\ref{IntAux1}) the parameter $t$ takes the values $0 \le
|\mbox{Re}\,(t+1)| < 1$ (see, e.g., Ref.~\refcite{RG}).  However, in
order to determine the expansion coefficients of the second sum in
Eq.~(\ref{VrIntExp}), one has to go to the point $t=-3$, which is
outside of this range. Nevertheless, it can be done by making use of
the analytic continuation of the left-hand side of
Eq.~(\ref{IntAux1}). This continuation is unique and it is defined
obviously by the right-hand side of Eq.~(\ref{IntAux1}) all over  the
complex $t$--plane except for the points $t=-2 N$, with~$N$ being a
natural number. Fortunately, we are not dealing with these values of
the parameter~$t$, and we can put
\begin{equation}
\label{IntCont}
\int\limits_{0}^{\infty} \left.\frac{\sin Q}{Q^s} (\ln Q)^n\, d Q =
v(n, t)\right|_{t=-s}, \quad s \neq 2 N\quad (N=1,2,3,...\,).
\end{equation}
It should be noted here that this analytical continuation plays the
role of regularization of the expansion coefficients in
Eq.~(\ref{VrIntExp}). It is also convenient to introduce the
notations
\begin{equation}
\label{uwDef}
u_{n} = \left.\frac{2}{\pi}\, v(n, t)\right|_{t=-1}, \qquad
\omega_{m} = -\left.\frac{2}{\pi}\, v(m, t)\right|_{t=-3}.
\end{equation}
A few first coefficients (\ref{uwDef}) have a quite simple form,
namely $u_0 = 1$,$\;$ $u_1 = - \gamma$,$\;$  $u_2 = \gamma^2 +
\pi^2/12$; and  $\omega_0 = 1/2$,$\;$ $\omega_1 = 3/4 - \gamma/2$,$\;$
$\omega_2 =  \gamma^2/2 - 3 \gamma/2 +\pi^2/24 + 7/4$, where  $\gamma
\simeq 0.57721...$ denotes the Euler's constant (see Ref.~\refcite{IJMPA}).
It is worth noting here that  the leading coefficients $u_0$, $\,\omega_0$,
and $\omega_1$ have also been calculated in Ref.~\refcite{PRD1}, but by
making use of another technique.

     Thus, the static quark--antiquark potential (\ref{VrIntExp})
can be represented now in the following form:
\begin{equation}
\label{VrAsympt}
V(r) \simeq \frac{8 \pi}{3 \beta_0}\, \frac{1}{r_0} \left[
\frac{1}{R} \sum_{n=0}^{n_0} \frac{u_n}{(\ln R)^{n+1}}
+ R \sum_{m=0}^{m_0} \frac{\omega_m}{(\ln R)^{m+1}}\right],
\qquad R=\frac{r}{r_0}.
\end{equation}
At small distances the potential (\ref{VrAsympt}) possesses the standard
behavior, determined by the asymptotic freedom
\begin{equation}
V(r) = \frac{8 \pi}{3 \beta_0}\, \frac{1}{r_0}\,
\frac{1}{R\, \ln R}, \qquad r \to 0.
\end{equation}
At the same time, it proves to be rising at large distances
\begin{equation}
V(r) = \frac{8 \pi}{3 \beta_0}\, \frac{1}{r_0}\,
\frac{R}{2\, \ln R}, \qquad r \to \infty,
\end{equation}
implying the confinement of quarks. It is of a particular interest to
mention here that a similar rising behavior of the \QbQ potential
has been proposed\cite{FLW} a long time ago proceeding from the
phenomenological assumptions.

     Equation (\ref{VrAsympt}) describes the behavior of the static
quark--antiquark potential~(\ref{VrInt}) at small and large
distances. However, its straightforward extrapolation to all
distances encounters poles of different orders at the point~$R=1$,
which apparently is an artifact of the expansion~(\ref{VrIntExp}).
For the practical purposes it would be undoubtedly useful to derive
an explicit expression for the \QbQ potential applicable for $0 < r <
\infty$.

     In order to construct an explicit interpolating formula for the
quark--antiquark potential~(\ref{VrInt}) we shall employ here the
following method. Let us modify the expansion (\ref{VrAsympt})
in a ``minimal'' way, by adding terms which only subtract the
singularities at the point $R=1$ and do not contribute to the derived
asymptotics. This leads to the following expression for the static
quark--antiquark potential
\begin{eqnarray}
V(r) &=& \frac{8 \pi}{3 \beta_0}\, \frac{1}{r_0} \left\{
\sum_{n=0}^{n_0} u_n
\left[\frac{1}{R\,(\ln R)^{n+1}} \right]_\Reg
+ \sum_{m=0}^{m_0} \omega_m
\left[ \frac{R}{(\ln R)^{m+1}}\right]_\Reg\right\},
\label{VrReg}
\end{eqnarray}
where $R=r/r_0$, and  at the leading orders $[1/(R\,\ln R)]_\Reg =
1/(R\,\ln R) - 1/P$, $[1/(R\,(\ln R)^2)]_\Reg = 1/(R\,(\ln R)^2) -
1/P^{2}$, and $[R/\ln R]_\Reg = R/\ln R - 1/P$, $[R/(\ln R)^2]_\Reg =
R/(\ln R)^2 - 2/P - 1/P^{2}$, $P=R-1$ (see also Ref.~\refcite{IJMPA}).

\begin{figure}[ht]
\centerline{\includegraphics[width=65mm]{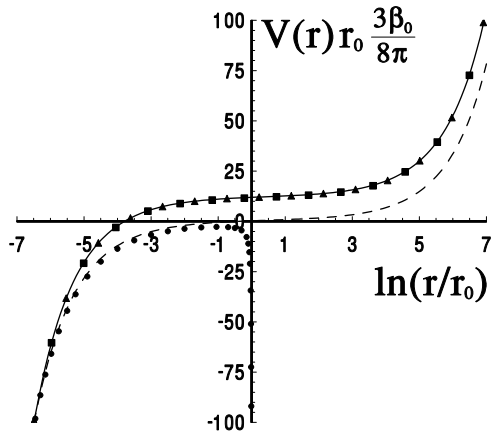}}
\caption{The quark--antiquark potential (\protect\ref{VrReg}) in
dimensionless units at different levels of approximation: $n_0 = m_0
= 0$ (dashed curve), $n_0 = m_0 = 4$ (solid curve), $n_0 = m_0 = 5$
($\blacktriangle$), and $n_0 = m_0 = 10$ ({\tiny $\blacksquare$}).
The one-loop perturbative result is shown by dotted curve.}
\label{Fig:VrAppr}
\end{figure}

     The numerical analysis of Eq.~(\ref{VrReg}) revealed
that for practical purposes it is enough to retain the first five
expansion terms ($n_0 = m_0 = 4$) therein. It turns out that the
potential~(\ref{VrReg}) itself and the  corresponding estimation of
the value of parameter $\LQCD$ are not affected by higher--order
contributions. In particular, the curves (\ref{VrReg}) for $n_0 = m_0
= 5$ and for  $n_0 = m_0 = 10$ are practically indistinguishable of
the curve corresponding to $n_0 = m_0 = 4$ over the whole region
$0<r<\infty$ (see Figure~\ref{Fig:VrAppr}). And the higher--order
estimations of the parameter $\LQCD$ vary within $0.5\,\%$ of the
value obtained at the $n_0 = m_0 = 4$ (see Sec.~4 further).

     Thus, we arrive at the following explicit expression for the
static quark--antiquark potential:
\begin{eqnarray}
V(r) &=& V_{0} + \frac{8 \pi}{3 \beta_0}\, \frac{1}{r_0} \Biggl\{
\frac{1}{R}\Biggl[
\frac{1}{\ln R} - \frac{0.577}{(\ln R)^2}
+ \frac{1.156}{(\ln R)^3}
- \frac{4.021}{(\ln R)^4} + \frac{15.018}{(\ln R)^5}\Biggr]
\nonumber \\ &&
+ R \Biggl[\frac{0.500}{\ln R} + \frac{0.461}{(\ln R)^2}
+ \frac{1.462}{(\ln R)^3}
+ \frac{3.185}{(\ln R)^4}
+ \frac{17.844}{(\ln R)^5}\Biggr]
\nonumber \\ &&
+ \frac{21.489}{1-R}
- \frac{62.484}{(1-R)^2}
+ \frac{98.694}{(1-R)^3} - \frac{84.143}{(1-R)^4}
+ \frac{32.861}{(1-R)^5} \Biggr\},
\label{Vr}
\end{eqnarray}
where $R=r/r_0$. In order to reproduce the correct short-distance
behavior of the $\mbox{Q}\overline{\mbox{Q}}$~potential, the
dimensional parameter~$r_0$ in this equation has to be identified
with $\Lambda_{\overline{\mbox{\tiny MS}}}$ by the relation
$r_0^{-1} = \Lambda \exp(\gamma)$ (see, e.g.,
Refs.~\refcite{Melles,Peter,Kiselev2}). It is straightforward to
verify that the potential (\ref{Vr}) satisfies also the concavity
condition
\begin{equation}
\frac{d\,V(r)}{d\,r} > 0, \qquad \frac{d^2 V(r)}{d\, r^2} \le 0,
\end{equation}
which is a general property of the gauge theories (see
Ref.~\refcite{SB} for the details).

\section{Discussion}

     In this section we are going to apply the obtained results to
the study of recent quenched lattice simulation data\cite{Bali2000}
on the static quark--antiquark potential. In particular, the value of
the parameter $\LQCD$ will be estimated by comparing the derived
expression for the \QbQ potential (\ref{Vr}) with these lattice data.

\begin{figure}[ht]
\centerline{\includegraphics[width=65mm]{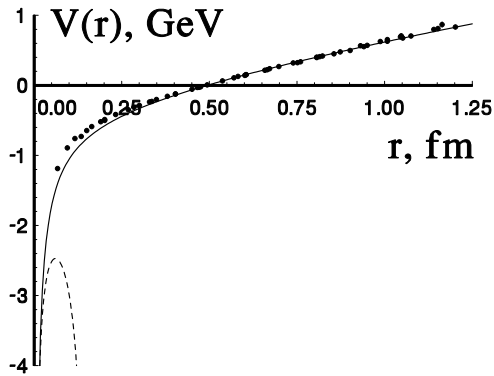}}
\caption{Comparison of the quark--antiquark potential $V(r)$ defined
by Eq.~(\protect\ref{Vr}) (solid curve) with the quenched lattice
simulation data\protect\cite{Bali2000} ($\bullet$). The values of the
parameters are: $\Lambda = 670\,$MeV, $\nfs = 0$,
$V_{0}=-3.164\,$GeV. The dashed curve corresponds to the relevant
one-loop perturbative result.
\label{Fig:Vr}}
\end{figure}

     Thus, a fit of the quark--antiquark potential (\ref{Vr}) to the
quenched lattice simulation data\cite{Bali2000} has been performed
with the use of the least square method. The varied parameters in
Eq.~(\ref{Vr}) were $r_0$ and the additive self--energy constant
$V_0$. The result of the fit is presented in Figure~\ref{Fig:Vr}.
This figure shows that in the nonperturbative physically--relevant
range $0.3\,\mbox{fm} \lesssim r \lesssim 1.2\,\mbox{fm}$, in which
the average quark separations $\sqrt{\langle r^2\rangle}$ for
quarkonia sits,\cite{Bali9799} the \QbQ potential (\ref{Vr})
reproduces the lattice data\cite{Bali2000} fairly well. At the same
time, in the region $r \lesssim 0.05\,\mbox{fm}$ the derived
potential coincides with the perturbative result\footnote{The
reliability of the perturbative quark--antiquark potential at
small distances has been discussed in Refs.~\refcite{Peter,BGT,HMP}.}
(dashed curve in
Fig.~\ref{Fig:Vr}). The difference between the lattice data and the
expression (\ref{Vr}) in the intermediate range $0.05\,\mbox{fm}
\lesssim r \lesssim 0.3\,\mbox{fm}$ may be explained by the presence
of additional nonperturbative contributions at these distances (see,
e.g., Refs.~\refcite{ShortString,Lee}). But the detailed investigation
of this matter is beyond the scope of the paper.

     Thus, the estimation of the parameter $\LQCD$ in the course of
this comparison gives $\Lambda^{\!(\nfs=0)} = (670 \pm 8)\,$MeV (this
value corresponds to the one-loop level with $\nf=0$ active quarks).
The uncertainty here has been calculated by making use of the
``$3 \sigma$--criterion'' (see Ref.~\refcite{Data}). It is worth
noting that the fit with the use of the maximum likelihood method
results in even better reproduction of the lattice data and
gives a similar value $\Lambda^{\!(\nfs=0)} \simeq 645\,$MeV.
Further, in order to collate the obtained estimation with the earlier
ones, one has to continue it to the region of three active
quarks. It was performed by employing the matching procedure, and
gives\footnote{It is interesting to note here that a similar values
of parameter $\LQCD$ have also been obtained within different
approaches to this issue (see Refs.~\refcite{LeTo,BGT,CeHe}).}  the
value $\Lambda^{\!(\nfs=3)} = (590 \pm 10)\,$MeV. The latter is in a
good agreement with all the previous estimations of this
parameter\cite{PRD2,ConfV} [$\Lambda^{\!(\nfs=3)} = (550 \pm
50)\,$MeV] in the framework of the approach in hand.

\section{Conclusion}

     In the paper the QCD analytic invariant charge has been applied
to  study of the quenched lattice simulation data on the static
quark--antiquark  potential. This strong running coupling enables one
to obtain explicitly the confining quark--antiquark potential in the
framework of the one-gluon exchange model. A technique for evaluation
the integrals of a specific form, developed in this paper, allows one
to examine the asymptotic behavior of the derived potential at small
and large distances with high accuracy. An explicit formula for the
quark--antiquark potential, which interpolates between these
asymptotics and satisfies the concavity condition, is obtained. The
derived potential has the standard form determined by asymptotic
freedom at small distances. At the same time, in the nonperturbative
physically--relevant region $0.3\,\mbox{fm}\lesssim r \lesssim
1.2\,\mbox{fm}$ this potential agrees fairly well with the lattice
simulation data. The value of the parameter $\LQCD$ is estimated for
the case of pure gluodynamics. It is found to be consistent with all
the previous estimations of $\LQCD$ in the framework of the approach
developed.

\section*{Acknowledgments}

     The author is grateful to Professor C.~Roiesnel and Professor
B.~Pire for the fruitful discussions and valuable advises, and to
Professor G.S.~Bali for supplying the relevant  lattice simulation
data and useful comments. The author also thanks the CPHT Ecole
Polytechnique for the very warm hospitality. The partial support of
RFBR (grants 02--01--00601 and 04--02--81025) and NSh--2339.2003.2
is appreciated.


\begin{thebibliography}{99}
\bibitem{Redmond} P.J.~Redmond, {\it Phys.\ Rev.} {\bf 112}, 1404 (1958);
P.J.~Redmond and J.L.~Uretsky, {\it Phys.\ Rev.\ Lett.} {\bf 1}, 147 (1958).

\bibitem{BLS} N.N.~Bogoliubov, A.A.~Logunov, and D.V.~Shirkov,
         {\it Zh.\ Exp.\ Teor.\ Fiz.} {\bf 37}, 805 (1959)
         [{\it Sov.\ Phys.\ JETP} {\bf 37}, 574 (1960)].

\bibitem{SS97} D.V.~Shirkov and I.L.~Solovtsov, {\it Phys.\ Rev.\ Lett.}
         {\bf 79}, 1209 (1997).

\bibitem{DV01} I.L.~Solovtsov and D.V.~Shirkov,
         {\it Teor.\ Mat.\ Fiz.} {\bf 120}, 482 (1999)
         [{\it Theor.\ Math.\ Phys.} {\bf 120}, 1220 (1999)];
         D.V.~Shirkov, {\it Eur.\ Phys.\ J.} {\bf C22}, 331 (2001).

\bibitem{PRD1} A.V.~Nesterenko, {\it Phys.\ Rev.} {\bf D62}, 094028 (2000).

\bibitem{PRD2} A.V.~Nesterenko, {\it Phys.\ Rev.} {\bf D64}, 116009 (2001).

\bibitem{IJMPA} A.V.~Nesterenko, {\it Int.\ J.\ Mod.\ Phys.} {\bf A18},
         5475 (2003).

\bibitem{MPLA1} A.V.~Nesterenko, {\it Mod.\ Phys.\ Lett.} {\bf A15}, 2401
         (2000).

\bibitem{MPLA2} A.V.~Nesterenko and I.L.~Solovtsov, {\it Mod.\ Phys.\ Lett.}
         {\bf A16}, 2517 (2001).

\bibitem{Bali2000} G.S.~Bali {\it et al.}, (SESAM and T$\chi$L
         Collaborations), {\it Phys.\ Rev.} {\bf D62}, 054503 (2000).

\bibitem{UKQCD} D.A.~Smith and M.J.~Teper (UKQCD Collaboration),
         {\it Phys.\ Rev.} {\bf D58}, 014505 (1998);
         A.~Ringwald and F.~Schrempp, {\it Phys.\ Lett.}
         {\bf B459}, 249 (1999);
         {\bf B503}, 331 (2001).

\bibitem{QCD03} A.V.~Nesterenko, in {\it Proceedings of the Tenth High Energy
         Physics International Conference in Quantum Chromodynamics,
         Montpellier, France, 2003} (to be published);
         arXiv: {\sf hep-ph/0307283}.

\bibitem{ConfV} A.V.~Nesterenko, in {\it Proceedings of the Fifth
         International Conference on Quark Confinement and the Hadron
         Spectrum, Gargnano, Italy, 2002}, edited by N.~Brambilla and
         G.~Prosperi (World Scientific, Singapore, 2003), p.~288;
         arXiv: {\sf hep-ph/0210122}.

\bibitem{Schrempp} F.~Schrempp, {\it J.\ Phys.} {\bf G28}, 915 (2002).

\bibitem{Ynd1} S.~Titard and F.J.~Yndurain, {\it Phys.\ Rev.} {\bf D49},
         6007 (1994).

\bibitem{Lucha91} W.~Lucha, F.F.~Schoberl, and D.~Gromes, {\it Phys.\ Rep.}
         {\bf 200}, 127 (1991).

\bibitem{Brambilla99} N.~Brambilla and A.~Vairo, arXiv: {\sf hep-ph/9904330}.

\bibitem{Kiselev1} V.V.~Kiselev and A.K.~Likhoded, {\it Usp.\ Fiz.\ Nauk}
         {\bf 172}, 497 (2002) [{\it Sov.\ Phys.\ Usp.} {\bf 45}, 455 (2002)].

\bibitem{LeTo} R.~Levine and Y.~Tomozawa, {\it Phys.\ Rev.} {\bf D19},
         1572 (1979).

\bibitem{RG} I.S.~Gradshteyn and I.M.~Ryzhik, {\it Table of Integrals,
         Series, and Products}, edited by A.~Jeffrey (Academic, London,
         1994).

\bibitem{FLW} G.~Fogleman, D.B.~Lichtenberg, and J.G.~Wills,
         {\it Lett.\ Nuovo Cimento} {\bf 26}, 369 (1979);
         D.B.~Lichtenberg and J.G.~Wills, {\it Nuovo Cimento}
         {\bf A47}, 483 (1978).

\bibitem{Melles} M.~Melles, {\it Phys.\ Rev.} {\bf D62}, 074019 (2000).

\bibitem{Peter} M.~Peter, {\it Phys.\ Rev.\ Lett.} {\bf 78}, 602 (1997);
         {\it Nucl.\ Phys.} {\bf B501}, 471 (1997).

\bibitem{Kiselev2} V.V.~Kiselev, A.E.~Kovalsky, and A.I.~Onishchenko,
         {\it Phys.\ Rev.} {\bf D64}, 054009 (2001);
         V.V.~Kiselev, A.K.~Likhoded, O.N.~Pakhomova, and V.A.~Saleev,
         {\it ibid.} {\bf D66}, 034030 (2002).

\bibitem{SB} E.~Seiler, {\it Phys.\ Rev.} {\bf D18}, 482 (1978);
             C.~Bachas, {\it ibid.} {\bf D33}, 2723 (1986).

\bibitem{Bali9799} G.S.~Bali, K.~Schilling, and A.~Wachter,
         {\it Phys.\ Rev.} {\bf D56}, 2566 (1997);
         G.S.~Bali and P.~Boyle, {\it ibid.} {\bf D59},
         114504 (1999).

\bibitem{BGT} W.~Buchmuller, G.~Grunberg, and S.-H.~H.~Tye, {\it Phys.\
         Rev.\ Lett.} {\bf 45}, 103 (1980);
         {\bf 45}, 587(E) (1980);
         W.~Buchmuller and S.-H.~H.~Tye, {\it Phys.\ Rev.} {\bf D24},
         132 (1981).

\bibitem{HMP} K.~Hagiwara, A.D.~Martin, and A.W.~Peacock,
         {\it Z.\ Phys.} {\bf C33}, 135 (1986).

\bibitem{ShortString} R.~Akhoury and V.I.~Zakharov, {\it Phys.\ Lett.}
         {\bf B438}, 165 (1998);
         G.S.~Bali, {\it ibid.} {\bf B460}, 170 (1999).

\bibitem{Lee} T.~Lee, {\it Phys.\ Rev.} {\bf D67}, 014020 (2003).

\bibitem{Data} S.L.~Meyer, {\it Data Analysis for Scientists and Engineers}
         (Wiley, New York, 1975).

\bibitem{CeHe} W.~Celmaster and F.S.~Henyey, {\it Phys.\ Rev.} {\bf D18},
         1688 (1978).

\end{thebibliography}
\end{document}